\documentclass[aps,prl,twocolumn,showpacs,superscriptaddress,reprint]{revtex4-1}
\usepackage{graphicx}
\usepackage{natbib}
\usepackage{amsmath}
\usepackage{amssymb}
\usepackage{bm}
\usepackage{color}
\usepackage{float}

\begin{document}

\title{Hybridization dynamics in CeCoIn$_5$ revealed by ultrafast optical spectroscopy }
\author{Y. P. Liu} 
\affiliation{State Key Laboratory of Electronic Thin Films and Integrated Devices, University of Electronic Science and Technology of China, Chengdu 610054, China}
\affiliation{Institute of Modern Physics, Fudan University, Shanghai 200433, China}
\author{Y. J. Zhang}
\affiliation{Center for Correlated Matter and Department of Physics, Zhejiang University, Hangzhou 310058, China}
\author{J. J. Dong}
\affiliation{Beijing National Laboratory for Condensed Matter Physics, Institute of Physics, Chinese Academy of Science, Beijing 100190, China}
\affiliation{University of Chinese Academy of Sciences, Beijing 100049, China}
\author{H. Lee} 
\affiliation{Center for Correlated Matter and Department of Physics, Zhejiang University, Hangzhou 310058, China}
\author{Z. X. Wei} 
\affiliation{State Key Laboratory of Electronic Thin Films and Integrated Devices, University of Electronic Science and Technology of China, Chengdu 610054, China}
\affiliation{Institute of Electronic and Information Engineering, University of Electronic Science and Technology of China, Dongguan 523808, China}
\author{W. L. Zhang} 
\affiliation{State Key Laboratory of Electronic Thin Films and Integrated Devices, University of Electronic Science and Technology of China, Chengdu 610054, China}
\author{C. Y. Chen}
\affiliation{Institute of Modern Physics, Fudan University, Shanghai 200433, China}
\author{H. Q. Yuan} 
\affiliation{Center for Correlated Matter and Department of Physics, Zhejiang University, Hangzhou 310058, China}
\author{Yi-feng Yang}
\email{yifeng@iphy.ac.cn}
\affiliation{Beijing National Laboratory for Condensed Matter Physics, Institute of Physics, Chinese Academy of Science, Beijing 100190, China}
\affiliation{University of Chinese Academy of Sciences, Beijing 100049, China}
\affiliation{Songshan Lake Materials Laboratory, Dongguan 523808, China}
\author{J. Qi}
\email{jbqi@uestc.edu.cn} 
\affiliation{State Key Laboratory of Electronic Thin Films and Integrated Devices, University of Electronic Science and Technology of China, Chengdu 610054, China}

\date{\today}

\begin{abstract}
We investigate the quasiparticle dynamics in the prototype heavy fermion CeCoIn$_5$ using ultrafast optical pump-probe spectroscopy. Our results indicate that this material system undergoes hybridization fluctuations before the establishment of heavy electron coherence, as the temperature decreases from $\sim$120 K ($T^\dagger$) to $\sim$55 K ($T^*$ ). We reveal that the anomalous coherent phonon softening and damping reduction below $T^*$ are directly associated with the emergence of collective hybridization. We also discover a distinct collective mode with an energy of $\sim$8 meV, which may be experimental evidence of the predicted unconventional density wave. Our findings provide key information for understanding the hybridization dynamics in heavy fermion systems. 
\end{abstract}
\maketitle

In rare-earth or actinide intermetallics, localized $f$ electrons can turn gradually into itinerant heavy electrons with lowering temperature, with an effective mass reaching hundreds times that of free electrons \cite{Coleman2007}. In transport measurement, the transition occurs typically below a common temperature, $T^*$, often called the coherence temperature \cite{Yang2008}. It is generally believed that $T^*$ marks the onset of collective hybridization between localized $f$ moments and conduction electrons, causing the emergence of heavy electrons at lower temperatures \cite{Yang2016}. However, recent angle-resolved photoemission spectroscopy (ARPES) measurements seem to indicate the presence of hybridization already at much higher temperatures and no peculiar signatures were observed across $T^*$ \cite{Chen_PRB_2017,Koitzsch_PRB_2008,Koitzsch_PRB_2009,Koitzsch_PRB_2013}. This leads to a puzzling contradiction of interpretation among different probes and prevents a consistent understanding of the heavy fermion physics.

To explore this issue, we take CeCoIn$_5$ as an example, which has attracted intensive attentions in past years as a prototype heavy fermion compound. Previous studies have mostly focused on its equilibrium or quasi-equilibrium properties such as unconventional superconductivity \cite{Petrovic_JPC_2001}, exotic electronic states \cite{Kenzelmann2008}, and magnetic quantum criticality  \cite{Zaum2011}. Its localized-to-itinerant transition occurs at $T^*=50\pm10\,$K, as marked by a resistivity peak separating the high temperature insulating-like regime due to incoherent Kondo scattering from a coherent metallic state at low temperatures \cite{Petrovic_JPC_2001}. Similar crossover has been found in many bulk measurements and ascribed to a common origin owing to the emergence of heavy electrons \cite{Yang2008}. This has been further confirmed in the scanning tunneling microscopy/spectroscopy investigations which revealed an unusual quantum critical $E/T$ scaling in the local conductance below 60 K \cite{Aynajian_PRL_2012}. On the other hand, recent ARPES experiments \cite{Koitzsch_PRB_2008,Koitzsch_PRB_2009,Koitzsch_PRB_2013,Chen_PRB_2017} reported signatures of hybridization well above 100 K, where $f$ electrons are believed to be still localized. Fourier transform infrared spectroscopy (FTIR) also revealed a direct gap emerging above $T^*$, although the exact onset temperatures differ in various experiments \cite{Singley_PRB_2002,Mena_PRB_2005,Burch_PRB_2007}. This raises the question concerning the difference between hybridization dynamics below and above $T^*$ and why heavy electron coherence seems to only appear below $T^*$. 

Theoretically, the hybridization physics can be described by a slave boson or hybridization field \cite{Coleman2007}. It is thus speculated that heavy electron emergence might be accompanied with certain type of collective  excitations. Unfortunately, a direct detection of such excitations has been missing. Only very few experiments have paid attention to bosonic excitations (mostly phonons) in CeCoIn$_5$ \cite{Martinho_PRB_2004,Bel_PRL_2004}. Raman measurements reported anomalous phonon response across $T^*$ \cite{Martinho_PRB_2004}, while the Seebeck and Nernst coefficients revealed intriguing anomalies at about 20 K \cite{Bel_PRL_2004}. It is not clear whether these findings are closely connected with the quasiparticle dynamics. 

To fill in this gap, we report ultrafast optical pump-probe measurements on CeCoIn$_5$. This technique has been widely applied in the studies of correlated materials \cite{Basov_RMP_2011,Ultrafast_review,Demsar_JPC_2006}. It provides a unique way to probe the dynamics of excited fermionic quasiparticles through their couplings to collective bosonic excitations, and thus allows us to detect the concurrent responses of fermionic and bosonic fields. In comparison with all previous measurements, we observed anomalous but quite different quasiparticle relaxation above and below $T^*$. While the relaxation rate shows a clear reduction starting at around $T^\dagger\approx120\,$K and continuing below $T^*\approx55\,$K, it becomes strongly fluence-dependent below $T^*$. We argue that the fluence-dependent relaxation indicates a nonlinear effect which can be ascribed to bimolecular recombination of excited quasiparticles across a narrow indirect hybridization gap associated with the formation of coherent heavy electrons on the Kondo lattice, while the fluence-independent relaxation implies the (indirect) gap closing above $T^*$ and should originate from the effect of precursor hybridization fluctuations. Following the gap opening below $T^*$, an unusual renormalization of the coherent phonon energy/damping is observed in our experiment, disclosing a weak but noticeable coupling between fermionic quasiparticles and coherent lattice dynamics. We also observed a prominent collective mode below $\sim$20 K, probably associated with some unconventional density wave. Our observations may thus help to reconcile the seeming ``contradiction" among previous measurements. 

In the pump-probe experiments, the ultrafast time-resolved differential reflectivity $\Delta R(t)/R$ was measured on high quality single crystal CeCoIn$_5$ at a center wavelength of 800 nm ($\sim$1.55 eV) using a Ti:sapphire femtosecond laser with a pulse width of $\sim$35 fs, taken from room temperature down to 5 K \cite{Wang_PRL_2016,Wang_PRB_2018,Qi_PRL_2013}. Figure \ref{fig:deltaR}(a) shows the measured signals up to room temperatures (see more experimental details in the Supplemental Material \cite{note}). Upon photoexcitation, the $\Delta R/R$ signal exhibits an instantaneous rise, succeeded by lateral relaxation processes. The time evolution of $\Delta R/R$ is dominated by the electron-electron (e-e) and electron-boson scattering processes. The boson can be phonons or other collective excitations \cite{Ultrafast_review,Wang_PRB_2018}. Surprisingly, the $\Delta R/R$ signals also display clear damped oscillations, which are superimposed on the non-oscillating background. 

We first focus on the non-oscillatory signals. At low temperatures, a strong fluence-dependent relaxation was observed in the short timescale $t<$2 ps, as evidently demonstrated in Fig. \ref{fig:deltaR}(b). By contrast, the dynamics for $t>$2 ps keeps nearly unchanged as the pump fluence varies. Note that the second rise with ps timescale, quite strong below $\sim$10 K\cite{note}, is probably associated with the coupling of quasiparticles with some bosonic excitations of electronic origins entangled with the nonthermal e-e scatterings, and has been observed in many correlated systems \cite{Ultrafast_review,Demsar_JPC_2006,Wang_PRB_2018,Kusar_PRB_2005,Hinton_PRL_2013,Vishik_PRB_2016}. For quantitative study of the quasiparticle relaxation, we fit the data below $\sim$2 ps with a single exponential formula, $\Delta R/R=Ae^{-\gamma t}$, where $A$ and $\gamma$ are the amplitude and decay rate, respectively. The fitting was performed only for the time after the maximal $\Delta R/R$ \cite{Vishik_PRB_2016}. The derived $\gamma$ is plotted in Fig. \ref{fig:deltaR}(c) as a function of temperature for different pump fluences. Within our experimental resolution, clear fluence-dependent behavior was found below a critical temperature of 55($\pm$5) K, which is roughly equal to $T^*$ reported in transport measurement \cite{Petrovic_JPC_2001,Yang2008}, indicating a close relationship between quasiparticle relaxation and the heavy electron coherence. As the temperature is increased from $\sim$55 K, $\gamma$ becomes fluence-independent and exhibits an anomaly at $T^\dagger\approx120\,$K, above which it saturates. Interestingly, this higher temperature scale resembles those observed in the FTIR and ARPES experiments \cite{Singley_PRB_2002,Burch_PRB_2007,Koitzsch_PRB_2008,Koitzsch_PRB_2009,Koitzsch_PRB_2013,Chen_PRB_2017}, where the  hybridization was suggested to appear.

\begin{figure}
	\includegraphics[width=9cm]{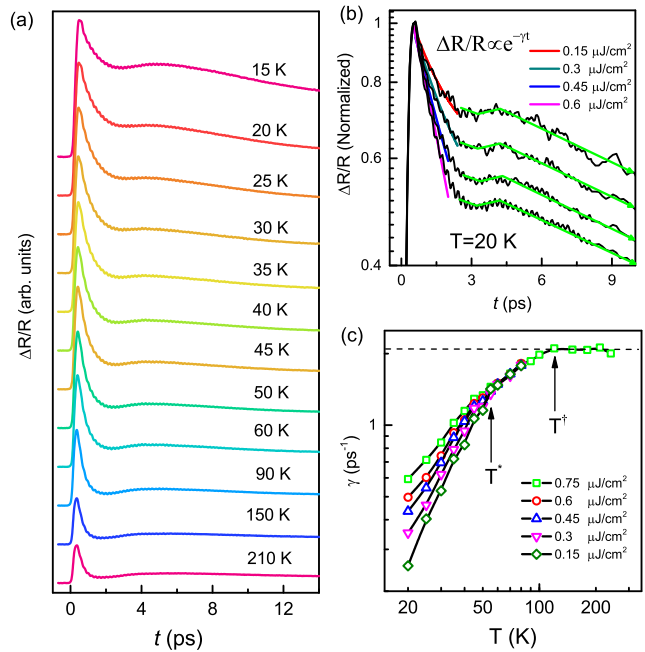}
	\caption{\label{fig:deltaR} (a) Typical $\Delta R(t)/R$ as a function of temperature at a pump fluence of $\sim$0.45 $\mu$J/cm$^2$. (b) $\Delta R(t)/R$ at 20 K as a function of the pump fluence. Relaxation below $\sim$2 ps shows strong fluence dependent. The relaxation can be fitted by a single exponential decay ($\propto e^{-\gamma t}$), where $\gamma$ is the decay rate. The fitting results below $\sim$2 ps are indicated by the solid lines with different colors. The green lines above $\sim$2 ps are guides to the eye. They are nearly the same except the shifting in $\Delta R/R$ axis. (c) The decay rate $\gamma$ as a function of temperature measured at different fluences. Two clear anomalies at $T^*$ and $T^\dagger$ are found ($T^*<T^\dagger$). Below $T^*$, $\gamma$ shows strong fluence-dependent. Above $T^\dagger$, $\gamma$ almost keeps constant.}
	\vspace*{-0.2cm}
\end{figure}

\begin{figure}
	\includegraphics[width=8.5cm]{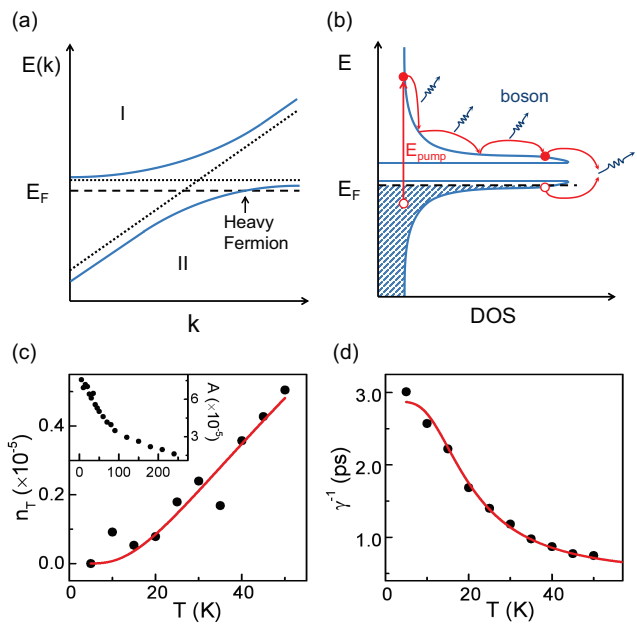}
	\caption{\label{fig:RTfit} (a) Illustration of the hybridization between local $f$ and conduction electrons below $T^*$, leading to the indirect gap, $\Delta_{\text{ind}}$, and the heavy fermion states near the Fermi energy $E_F$. (b) High energy excitations using 800 nm optical pulses in excess of $\Delta_{\text{ind}}$. Quasiparticles via high energy excitations decay to the gap edge via emission of high frequency phonons or other bosonic excitations. Subsequently, the bimolecular recombination dominates the decay and the relaxation rates become fluence dependent. (c) The density of thermally excited quasiparticles $n_T$ as a function of temperature below $T^*$. The inset shows the temperature dependence of the amplitude $A$. (d) Decay time $\gamma^{-1}$ as a function of temperature below $T^*$. The red lines are the fit using the RT model.}
	\vspace*{-0.2cm}
\end{figure}

The fluence-dependence of the decay rate below $T^*$ indicates a nonlinear effect of quasiparticle relaxation and may be understood from the well-known Rothwarf-Taylor (RT) model \cite{Rothwarf_PRL_1967}, which describes the time evolution of densities of coupled quasiparticles ($n$) and bosons ($N$). If there is a narrow energy gap ($\Delta_{\text{ind}}$) in the electron density of states (DOS), the decay of excited quasiparticles with energies larger than the gap will be governed by the emission of high frequency bosons that can subsequently re-excite electron-hole pairs.  A bimolecular recombination term then will dominate the quasiparticle relaxation when the recombination rate $R$ or the boson relaxation time is large, causing a nonlinear $Rn^2$ contribution and hence the strong fluence-dependence of $\gamma$. A schematic plot of this process is shown in Figs.~\ref{fig:RTfit}(a) and \ref{fig:RTfit}(b).

The RT model has been successfully applied to many correlated systems \cite{Kabanov_PRB_1999,Segre_2002_PRL,Demsar_JPC_2006,Vishik_PRB_2016}. The temperature dependence of $\gamma(T)$ and $A(T)$ can be used to elucidate quantitatively the gap formation \cite{Kabanov_PRB_1999,Demsar_JPC_2006,Chia_PRL_2007},
\begin{eqnarray}
\gamma(T)&\propto&\left[\frac{\delta}{\zeta n_T+1}+2n_T\right]\left(\Delta_{\text{ind}}+\alpha T\Delta_{\text{ind}}^4\right),\nonumber\\
n_T(T)&=&\frac{A(0)}{A(T)}-1\propto \left(T\Delta_{\text{ind}}\right)^p\text{e}^{-\Delta_{\text{ind}}/T},
\label{eq:RT-model1}
\end{eqnarray}
where $n_T$ are the density of quasiparticles thermally excited across the gap, $\alpha$, $\zeta$ and $\delta$ are fitting parameters and the value of $p$ ($0<p<1$) depends on the shape of the gapped DOS. Figures~\ref{fig:RTfit}(c) and \ref{fig:RTfit}(d) present a good fit to the experimental data below $T^*$, yielding an energy gap of $2\Delta_{\text{ind}}\approx8\,$meV, with $p=0.5$ from a typical Bardeen-Cooper-Schrieffer (BCS) form of the DOS  \cite{Demsar_JPC_2006}. In heavy fermion systems, this represents an indirect hybridization gap that opens only below $T^*$. Clearly, this gap is much smaller than the direct hybridization gap ($2\Delta_{\text{dir}}\approx75\,$meV) observed in FTIR experiments emerging above 100 K \cite{Singley_PRB_2002,Coleman_Nat_2005}. Theoretically, these two gaps should be roughly related by \cite{Okamura2007,Lonzarich2017}, $\Delta_{\text{dir}}\sim\sqrt{\Delta_{\text{ind}}W}$, where $W$ is the conduction bandwidth. Taking $\Delta_{\text{ind}}=4\,$meV and assuming $W=\pi^2k_B^2/3\gamma\approx 0.31\,$eV with $\gamma=7.6\,$mJ/mol K$^2$ from LaCoIn$_5$ \cite{Yang2008}, we obtain $\sqrt{\Delta_{\text{ind}}W}\approx 35\,$meV, in good agreement with the reported value of $\Delta_{\text{dir}}$. 

What happens above $T^*$? Obviously, the absence of the nonlinear effect indicates that the indirect hybridization gap is closed. In the mean-field theory, this takes place when the static hybridization becomes zero. However, precursor hybridization fluctuations should exist, affect the quasiparticle relaxation and cause the anomalous reduction in its decay rate until the temperature is further raised to $T^\dagger$. We should note that the constant $\gamma$ above $T^\dagger$ is also peculiar and cannot be described by the conventional two-temperature model \cite{Allen_PRL_1987}. Rather, it indicates a non-thermal relaxation via e-e collisions comparable with electron-boson scatterings  \cite{Kabanov_PRB_2008,Gadermaier_PRL_2010}. On such a time scale, the thermal distribution by e-e scatterings cannot be attained, even though the excited fermionic quasiparticles may relax close to the Fermi level. Possible candidates of bosonic excitations above $T^\dagger$ include phonons or spin fluctuations of localized $f$ moments. It is conceivable that the onset of precursor hybridization fluctuations tends to couple the non-equilibrium electrons near $E_F$ with fluctuating $f$ moments, suppress the e-e scatterings, and hence diminish $\gamma$ below $T^\dagger$. The rapid reduction of $\gamma$ below $T^\dagger$ indicates a fast growth of the hybridization fluctuations with lowering temperature.

\begin{figure}
	\includegraphics[width=9.2cm]{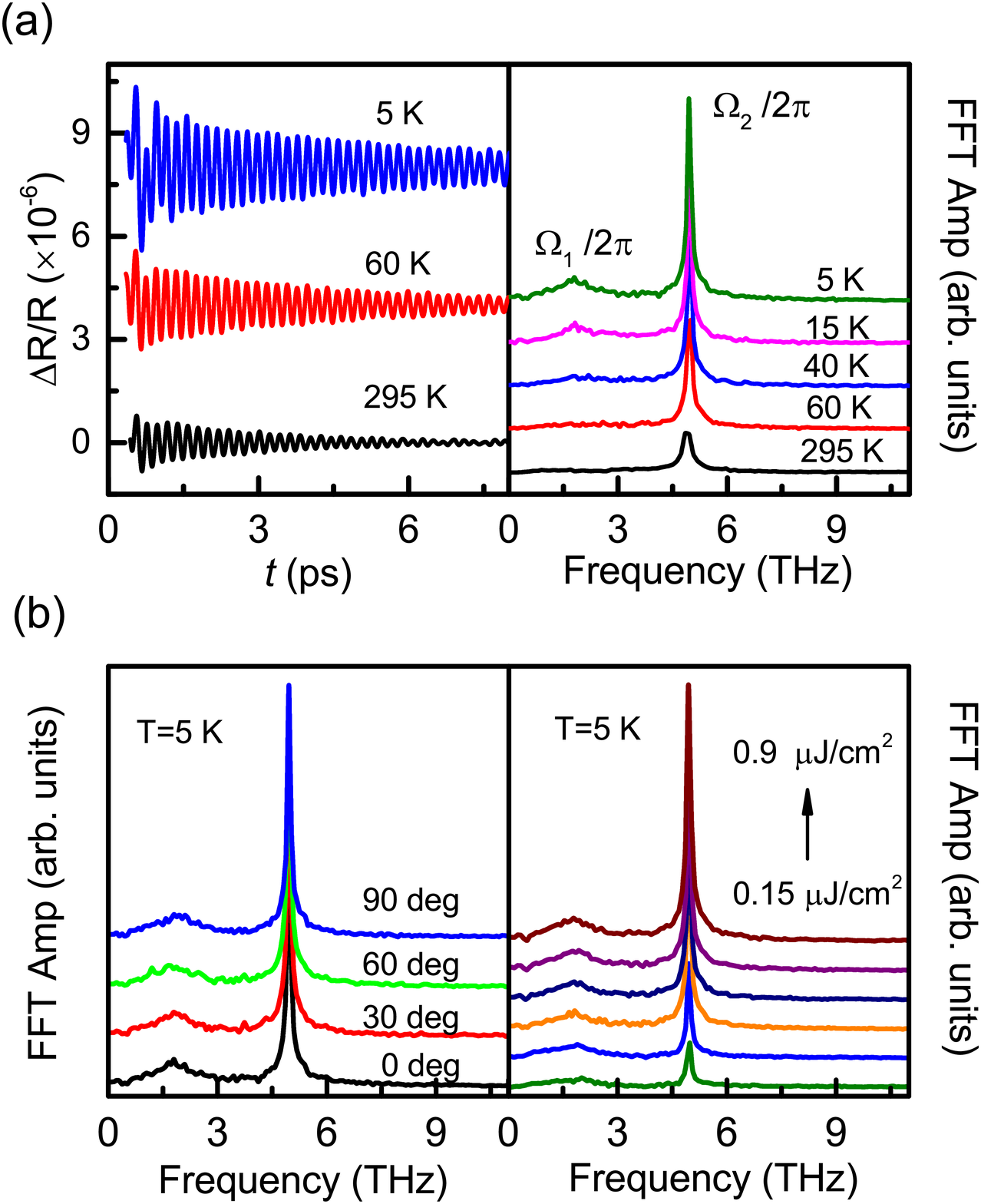}
	\caption{\label{fig:oscillation} (a) Extracted oscillations for two typical temperatures: 15 K and 290 K. Left panel includes the time-domain spectra, while the right is the corresponding Fast Fourier transform (FFT) frequency-domain data. Additional FFT data for oscillation at 20 K is also shown for comparison. (b) Oscillation modes as functions of pump polarization and pump fluence at 5 K. Curves are shifted for clarity. The polarization is defined by the angle with respect to the original position (pump-probe cross polarized) after rotating anticlockwise.The fluence increasing step is $\sim$0.15 $\mu$J/cm$^2$.}
	\vspace*{-0.2cm}
\end{figure}

The above analyses suggest a two-stage scenario for the hybridization dynamics. While $T^\dagger$ represents the onset of precursor hybridization fluctuations, $T^*$ marks its further development into a coherent heavy electron state protected by a tiny indirect hybridization gap. The latter is further manifested by the oscillation in $\Delta R(t)/R$, which typically arises from collective excitations such as coherent phonons and charge density wave during quasiparticle relaxation \cite{Tomeljak_PRL_2009,Chen_PRL_2017,Burch_PRL_2008,Qi_PRL_2013}. The oscillations with terahertz (THz) frequency are similar to those in some other systems \cite{Burch_PRL_2008,Qi_PRL_2013}. Specifically, the oscillatory components persisting up to room temperature are normally attributed to coherent phonons, which are initiated either via displacive excitations \cite{Zeiger_PRB_1992} or a photoexcitation-induced Raman process \cite{Merlin_SSS_1997}. Figure~\ref{fig:oscillation}(a) shows the oscillations in time and frequency domains at several typical temperatures, where two distinct high frequency modes were observed, i.e. $\Omega_1/2\pi\sim$2 THz and $\Omega_2/2\pi\sim$5 THz. However, only the latter survives up to room temperature, while the former is much weaker and quite prominent below $\sim$20 K. To extract their properties, we fit the oscillation pattern using the expression, 
\begin{equation}
 (\Delta R/R)_{\text{osc}}=\sum_{j=1,2}A_je^{-\Gamma_j t}\text{sin}(\Omega_j t+\phi_j),
\\
\label{eq:oscillation1}
\end{equation}
where $A_j$, $\Gamma_j$, $\phi_j$, and $\Omega_j$ are the amplitude, damping rate, phase, and frequency, respectively. $\Omega_j$ and $\Gamma_j$ are related for an underdamped harmonic oscillator, $\Omega_j=\sqrt{\omega_j^2-\Gamma_j^2}$, where $\omega_j$ is the natural frequency. 

Within our experimental resolution, the energy of $\omega_1$ mode does not depends on the temperature. Its prominence below $\sim$20 K accords well with previously observed anomalies in the Seebeck and Nernst coefficients \cite{Bel_PRL_2004}, which have been interpreted as an indication of unconventional density wave (UDW) \cite{Dora_PRB_2005}. Such UDW, however, has never been revealed by further experiments. Our observation of the $\omega_1$ collective mode seems to provide a plausible evidence for its existence. To elucidate its properties, we carried out further measurements by changing the polarization and fluence of the pump light. As shown in Fig.~\ref{fig:oscillation}(b), its independence on the polarization indicates that this mode is not associated with the asymmetric lattice vibrations \cite{Merlin_SSS_1997}, namely, the $E_g$ phonon, and may thus be fully symmetric. Fluence-dependent results show that its frequency and linewidth are nearly independent on the fluence, while its amplitude almost increases linearly as the fluence increases. The characterizing parameters of $\omega_1$ mode for the polarization- and fluence-dependent measurements can be found in the Supplemental Material~\cite{note}. Similar behavior was observed in the amplitude mode of the collective density wave excitations \cite{Chen_PRL_2017,Torchinsky_NMat_2013}. But there is still no explicit evidence to prove if it is associated with the spin \cite{Spin_resonance} or charge degree of freedom. Nonetheless, its energy scale of 2 THz ($\simeq8\,$meV) is very close to the indirect hybridization gap (2$\Delta_{\text{ind}}$), indicating a potential intimacy that could be a benchmark for future elaborate investigations.              

\begin{figure}
	\includegraphics[width=9cm]{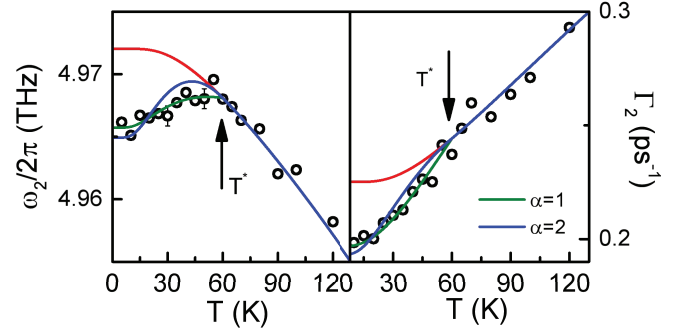}
	\caption{\label{fig:oscillationfit} The derived $\omega_2$ and $\gamma_2$ as a function of temperature using Eq.~(\ref{eq:oscillation1}), where the red lines represent the fit using the anharmonic phonon model, while the blue and green lines are the fit taking into consideration the contribution of Kondo singlets with different $\alpha$, as described in the main text.}
	\vspace*{-0.2cm}
\end{figure}

The $\omega_2$ mode can be identified as the coherent $A_{1g}$ phonon according to previous Raman measurement \cite{Martinho_PRB_2004}. Note that we did not observe the Raman mode with frequency of $\sim$1 THz. It could be that this mode is associated with the asymmetric $E_g$ phonon that decays extremely fast in the time domain \cite{Mahony_PRL_2019}. The extracted temperature evolutions of $\omega_2$ and $\Gamma_2$ are plotted in Fig.~\ref{fig:oscillationfit}. Instead of a monotonic increase of $\omega_2$ with lowering temperature, we see a sharp downturn at low temperatures. The temperature dependence of $\omega_2$ and $\Gamma_2$ above $T^*$ can be well explained by the anharmonic effect of optical phonons \cite{Balkanski_PRB_1983, Menendez_PRB_1984}, which typically includes contributions from lattice thermal expansion (Gr\"{u}neisen law) and anharmonic phonon-phonon coupling \cite{Qi_PRL_2013,Menendez_PRB_1984,Balkanski_PRB_1983,Hase_JPSJ_2015} (see fitting details in the Supplemental Material \cite{note}). However, both quantities were predicted to saturate at lower temperatures, inconsistent with our observations. Within the experimental resolution, such deviations take place in exact accordance with the coherence temperature $T^*$, as manifested in our analysis of the fluence-dependent quasiparticle relaxation. 

For quantitative understanding of the anomalous behavior below $T^*$, we assume the deviations $\delta\omega_2$ and $\delta\Gamma_2$ with respect to the expected anharmonic phonon contributions are proportional to $\langle b_i\rangle^\alpha$ with $\alpha=2$ as proposed in previous literature \cite{Burch_PRL_2008}. Here, 
$\langle b_i\rangle\propto [1-n_T(T)/n_T(T^*)]$ represents the density of Kondo singlets estimated from that of the quasiparticles ($n_T$) excited across the narrow gap. We then fit the experimental data using this assumption. However, the quality of the fit is not satisfactory (see Fig.~\ref{fig:oscillationfit}). Rather, a best fit with the same formula yields $\alpha=0.95\pm0.15$. For comparison, we also plot the fit with $\alpha=1$ in Fig.~\ref{fig:oscillationfit}(b). The excellent agreement with $\alpha=1$ suggests that $\delta\omega_2\propto \langle b_i\rangle$ and $\delta\Gamma_2\propto \langle b_i\rangle$. An alternate explanation is therefore needed in order to explain the phonon softening below $T^*$. Since $\langle b_i\rangle$ is directly associated with the indirect hybridization gap below $T^*$ in the mean-field theory \cite{Coleman2007}, it could be that the gap in the DOS constraints the electron scattering near the Fermi energy and as a consequence reduces the energy and damping of the phonons. We note that despite of the small effect of phonon renormalization, it still provides a useful probe of the collective hybridization. These findings are supported by our measurements in LaCoIn$_5$ \cite{note}, where no coherent heavy electron states exist.
 
Altogether, our observations not only confirm that the pump-probe technique allows for detection of the hybridization dynamics in a broad correlation time or length scale through quasiparticle relaxation, but also put the seemingly controversial ARPES and transport observations into the same unified framework. Though ARPES is capable of detecting the band bending at certain wave vectors near the Fermi surfaces caused by the onset of hybridization fluctuations at $T^\dagger$, it is greatly limited by its energy resolution, and so far unable to reveal the formation of the indirect hybridization gap with an order of meV below $T^*$. On the other hand, based on the resistivity data it is difficult to evidently reveal the initial fluctuations at $T^\dagger$ with a short correlation length occurring in a limited {\bf \textit{k}} space. Noticeable effect in resistivity is usually observed when a long-range coherence starts to build up below $T^*$ following opening of the indirect hybridization gap. This gap appears in the DOS and plays the role of protecting the composite heavy electron state so that $T^*$ marks the true coherence temperature of the Kondo lattice. Similar separation of the correlation lengths were also observed in recent experiments studying the charge density wave of LaTe$_3$ \cite{Zong_NPhys_2019}. It is thus conceivable that the heavy fermion physics is fundamentally associated with the temperature evolution of the correlation time or length scale of the hybridization fluctuations, and the heavy electron coherence only develops after a collective inter-site hybridization correlation is established on the Kondo lattice. This, however, is very different from the single-impurity Kondo physics, and so far has not been well studied and formulated in any current theory \cite{Yang2017,Hu2019}. The pump-probe experiments here provide critical information for the potential development of a microscopic understanding of the heavy fermion physics in the near future.      
	
We acknowledge valuable discussion from D. L. Feng and M. C. Wang. This work was supported by the National Natural Science Foundation of China (Grants Nos. 11974070, 11734006, 11774401, 11522435, and U1632275), the Frontier Science Project of Dongguan (2019622101004), the National Key R\&D Program of China (Grant Nos. 2017YFA0303100 and 2016YFA0300202), the Science Challenge Project of China (No. TZ2016004), and the CAS Interdisciplinary Innovation Team.


\begin{thebibliography}{text}
\bibitem{Coleman2007} P. Coleman, {\sl Heavy Fermions: Electrons at the Edge of Magnetism}, Handbook of Magnetism and Advanced Magnetic Materials Vol. 1 (Wiley, New York, 2007), pp. 95-148.
\bibitem{Yang2008} Y.-F. Yang, Z. Fisk, H.-O. Lee, J. Thompson, and D. Pines, Nature \textbf{454}, 611 (2008).
\bibitem{Yang2016} Y.-F. Yang, Rep. Prog. Phys. \textbf{79}, 074501 (2016).
\bibitem{Chen_PRB_2017} Q. Y. Chen, D. F. Xu, X. H. Niu, J. Jiang, R. Peng, H. C. Xu, C. H. P. Wen, Z. F. Ding, K. Huang, L. Shu, Y. J. Zhang, H. Lee, V. N. Strocov, M. Shi, F. Bisti, T. Schmitt, Y. B. Huang, P. Dudin, X. C. Lai, S. Kirchner, H. Q. Yuan, and D. L. Feng, Phys. Rev. B {\bf 96}, 045107 (2017).
\bibitem{Koitzsch_PRB_2008} A. Koitzsch, S. V. Borisenko, D. Inosov, J. Geck, V. B. Zabolotnyy, H. Shiozawa, M. Knupfer, J. Fink, B. B\"uchner, E. D. Bauer, J. L. Sarrao, and R. Follath, Phys. Rev. B {\bf 77}, 155128 (2008).
\bibitem{Koitzsch_PRB_2009} A. Koitzsch, I. Opahle, S. Elgazzar, S. V. Borisenko, J. Geck, V. B. Zabolotnyy, D. Inosov, H. Shiozawa, M. Richter, M. Knupfer, J. Fink, B. B\"uchner, E. D. Bauer, J. L. Sarrao, and R. Follath, Phys. Rev. B {\bf 79},075104 (2009).
\bibitem{Koitzsch_PRB_2013} A. Koitzsch, T. K. Kim, U. Treske, M. Knupfer, B. B\"uchner, M. Richter, I. Opahle, R. Follath, E. D. Bauer, and J. L. Sarrao, Phys. Rev. B {\bf 88}, 035124 (2013).
	
\bibitem{Petrovic_JPC_2001} C. Petrovic, P. G. Pagliuso, M. F. Hundley, R. Movshovich, J. L. Sarrao, J. D. Thompson, Z. Fisk, and P. Monthoux, J. Phys. Condens. Matter {\bf 13}, L337 (2001).
\bibitem{Kenzelmann2008} M. Kenzelmann, Th. Str\"assle, C. Niedermayer, M. Sigrist, B. Padmanabhan, M. Zolliker, A. D. Bianchi, R. Movshovich, E. D. Bauer, J. L. Sarrao, and J. D. Thompson, Science {\bf 321}, 1652 (2008)
\bibitem{Zaum2011} S. Zaum, K. Grube, R. Sch\"afer, E. D. Bauer, J. D. Thompson, and H. v. L\"ohneysen, Phys. Rev. Lett. {\bf 106}, 087003 (2011).

\bibitem{Aynajian_PRL_2012} P. Aynajian, E. H. da Silva Neto, A. Gyenis, R. E. Baumbach, J. D. Thompson, Z. Fisk, E. D. Bauer, and A. Yazdani, Nature {\bf 486}, 201 (2012).
\bibitem{Singley_PRB_2002} E. J. Singley, D. N. Basov, E.D. Bauer, and M. B. Maple, Phys. Rev. B {\bf 65}, 161101 (2002).
\bibitem{Mena_PRB_2005} F. P. Mena, D. van der Marel, and J. L. Sarrao, Phys. Rev. B {\bf 72}, 045119 (2005).
\bibitem{Burch_PRB_2007} K. S. Burch, S. V. Dordevic, F. P. Mena, A. B. Kuzmenko, D. van der Marel, J. L. Sarrao, J. R. Jeffries, E. D. Bauer, M. B. Maple, and D. N. Basov, Phys. Rev. B {\bf 75}, 054523 (2007).


\bibitem{Martinho_PRB_2004} H. Martinho, P. G. Pagliuso, V. Fritsch, N. O. Moreno, J. L. Sarrao, and C. Rettori, Phys. Rev. B {\bf 75}, 045108 (2007).

\bibitem{Bel_PRL_2004} R. Bel, K. Behnia, Y. Nakajima, K. Izawa, Y. Matsuda, H. Shishido, R. Settai, and Y. Onuki, Phys. Rev. Lett. {\bf 92}, 217002 (2004).

\bibitem{Basov_RMP_2011} D. N. Basov, R. D. Averitt, D. van der Marel, M. Dressel, and K. Haule, Rev. Mod. Phys. {\bf 83}, 471 (2011).
\bibitem{Ultrafast_review} C. Giannetti, M. Capone, D. Fausti, M. Fabrizio, F. Parmigiani, D. Mihailovic, Adv. Phys. {\bf 65}, 58-238 (2016).
\bibitem{Demsar_JPC_2006} J. Demsar, J. L. Sarrao, and A. J. Taylor, J. Phys. Condens. Matter \textbf{18}, R281 (2006).

\bibitem{Wang_PRL_2016} M. C. Wang, S. Qiao, Z. Jiang, S. N. Luo, J. Qi, Phys. Rev. Lett. {\bf 116}, 036601 (2016).
\bibitem{Wang_PRB_2018} M. C. Wang, H. S. Yu, J. Xiong, Y.-F. Yang, S. N. Luo, K. Jin, and J. Qi, Phys. Rev. B \textbf{97}, 155157 (2018). 
\bibitem{Qi_PRL_2013} J. Qi, T. Durakiewicz, S. A. Trugman, J.-X. Zhu, P. S. Riseborough, R. Baumbach, E. D. Bauer, K. Gofryk, J.-Q. Meng, J. J. Joyce, A. J. Taylor, R. P. Prasankumar, Phys. Rev. Lett. {\bf 111}, 057402 (2013).
\bibitem{note}  See Supplemental Material for detailed experimental setup and additional data of CeCoIn$_5$ and LaCoIn$_5$, which includes Refs.\cite{Takeuchi_JPCM_2002,Hilton_PRL_2002}.

\bibitem{Takeuchi_JPCM_2002} T. Takeuchi, H. Shishido, S. Ikeda, R. Settai, Y. Haga and Y. Onuki, J. Phys. Cond Matt. {\bf14}, L261(2002).

\bibitem{Hilton_PRL_2002} D. J. Hilton, C. L. Tang, Phys. Rev. Lett. {\bf89}, 146601(2002).

\bibitem{Kusar_PRB_2005} P. Kusar, J. Demsar, D. Mihailovic, S. Sugai, Phys. Rev. B {\bf 72}, 014544 (2005). 
\bibitem{Hinton_PRL_2013} J. P. Hinton, J. D. Koralek, G. Yu, E. M. Motoyama, Y. Lu, A. Vishwanath, M. Greven, J. Orenstein, Phys. Rev. Lett. {\bf 110}, 217002 (2013).
\bibitem{Vishik_PRB_2016} I. M. Vishik, F. Mahmood, Z. Alpichshev, N. Gedik, J. Higgins, and R. L. Greene, Phys. Rev. B {\bf 95}, 115125 (2017).
\bibitem{Rothwarf_PRL_1967} A. Rothwarf and B. N. Taylor, Phys. Rev. Lett. 19, 27 (1967).

\bibitem{Kabanov_PRB_1999} V. V. Kabanov, J. Demsar, B. Podobnik, and D. Mihailovic, Phys. Rev. B {\bf 59}, 1497 (1999).
\bibitem{Segre_2002_PRL} G. P. Segre, N. Gedik, J. Orenstein, D. A. Bonn, R. Liang, and W. N. Hardy, Phys. Rev. Lett. 88, 137001 (2002).
\bibitem{Chia_PRL_2007} E. E. M. Chia, J.-X. Zhu, D. Talbayev, R. D. Averitt, A. J. Taylor, K.-H. Oh, I.-S. Jo, and S.-I. Lee, Phys. Rev. Lett. {\bf 99}, 147008 (2007).
\bibitem{Coleman_Nat_2005} P. Coleman and A. J. Schofield, Nature {\bf 433}, 226 (2005).
\bibitem{Okamura2007} 
H. Okamura, T. Watanabe, M. Matsunami, T. Nishihara, N. Tsujii, T. Ebihara, H. Sugawara, H. Sato, Y. \-Onuki, Y. Isikawa, T. Takabatake, and T. Nanba, J. Phys. Soc. Jpn. {\bf 76}, 023703 (2007).
\bibitem{Lonzarich2017} G. Lonzarich, D. Pines, and Y.-F. Yang, Rep. Prog. Phys. \textbf{80}, 024501 (2017).


\bibitem{Allen_PRL_1987} P. B. Allen, Phys. Rev. Lett. 59, 1460 (1987).
\bibitem{Kabanov_PRB_2008} V. V. Kabanov and A. S. Alexandrov, Phys. Rev. B {\bf 78}, 174514 (2008).
\bibitem{Gadermaier_PRL_2010} C. Gadermaier, A. S. Alexandrov, V.V. Kabanov, P. Kusar, T. Mertelj, X. Yao, C. Manzoni, D. Brida, G. Cerullo, and D. Mihailovic, Phys. Rev. Lett. {\bf 105}, 257001 (2010).

\bibitem{Burch_PRL_2008} K. S. Burch, Elbert E. M. Chia, D. Talbayev, B. C. Sales, D. Mandrus, A. J. Taylor, and R. D. Averitt, Phys. Rev. Lett. {\bf 100}, 026409 (2008).

\bibitem{Tomeljak_PRL_2009} A. Tomeljak, H. Sch\"afer, D. St\"adter, M. Beyer, K. Biljakovic, and J. Demsar, Phys. Rev. Lett. {\bf 102}, 066404 (2009).
\bibitem{Chen_PRL_2017} R. Y. Chen, S. J. Zhang, M. Y. Zhang, T. Dong, and N. L. Wang, Phys. Rev. Lett. {\bf 118}, 107402 (2017).
\bibitem{Torchinsky_NMat_2013} D. H. Torchinsky, F. Mahmood, A. T. Bollinger, I. Božović and Nuh Gedik, Nat. Mater. {\bf 12}, 387 (2013).
\bibitem{Spin_resonance} No spin/magnetic excitations were reported near 2 THz. See references: J. Panarin, S. Raymond, G. Lapertot, and J. Flouquet, J. Phys. Soc. Jpn. {\bf 78}, 113706 (2009); S. Raymond and G. Lapertot, Phys. Rev. Lett. {\bf 115}, 037001 (2015).

\bibitem{Zeiger_PRB_1992} H. J. Zeiger, J. Vidal, T. K. Cheng, E. P. Ippen, G. Dresselhaus, and M. S. Dresselhaus, Phys. Rev. B {\bf 45}, 768 (1992).
\bibitem{Merlin_SSS_1997} R. Merlin, Solid State Commun. {\bf 102}, 207 (1997).
\bibitem{Dora_PRB_2005} B. D\'ora, K. Maki, A. Virosztek, and A. V\'anyolos, Phys. Rev. B {\bf 71}, 172502 (2005).
\bibitem{Mahony_PRL_2019} S. M. O'Mahony, F. Murphy-Armando, \'E. D. Murray, J. D. Querales-Flores, I. Savi\'c, and S. Fahy, Phys. Rev. Lett. {\bf 123}, 087401 (2019).

\bibitem{Balkanski_PRB_1983} M. Balkanski, R. F. Wallis, and E. Haro, Phy. Rev. B \textbf{28}, 1928 (1983).
\bibitem{Menendez_PRB_1984} J. Menendez and M. Cardona, Phys. Rev. B {\bf 29}, 2051 (1984).
\bibitem{Hase_JPSJ_2015} M. Hase, K. Ushida, and M. Kitajima, J. Phys. Soc. Jpn. \textbf{84}, 024708 (2015).

\bibitem{Zong_NPhys_2019} A. Zong, A. Kogar, Y.-Q. Bie, T. Rohwer, C. Lee, E. Baldini, E. Erge\c{c}en, M. B. Yilmaz, B. Freelon, E. J. Sie, H. Zhou, J. Straquadine, P. Walmsley, P. E. Dolgirev, A. V. Rozhkov, I. R. Fisher, P. Jarillo-Herrero, B. V. Fine, and N. Gedik, Nat. Phys. {\bf 15}, 27 (2019).

\bibitem{Yang2017} Y.-F. Yang, D. Pines, and G. Lonzarich, Proc. Natl. Acad. Sci. USA \textbf{114}, 6250 (2017).
\bibitem{Hu2019} D. Hu, J.-J. Dong, and Y.-F. Yang, Phys. Rev. B { \bf 100}, 195133 (2019).



\end{thebibliography}
\end{document}